\documentstyle[12pt]{article}

\newcommand {\vs}[1]  { \vspace*{#1 cm} }

\newcounter{eq}
\newcounter{sc}

\newcommand {\AP}   {Ann. of Phys.}
\newcommand {\CQG}  {Class. Quantum. Grav.}

\newcommand {\IJMP}  {Int. J. Mod. Phys.}

\newcommand {\MPL}  {Mod. Phys. Lett.}
\newcommand {\NP}   {Nucl. Phys.}

\newcommand {\PR}   {Phys. Rev.}

\def\overleftrightarrow#1{\vbox{\ialign{##\crcr
 $\leftrightarrow$\crcr\noalign{\kern-1pt\nointerlineskip}
 $\hfil\displaystyle{#1}\hfil$\crcr}}}

\newlength{\minitwocolumn}
\begin{document}

\begin{flushright}
DPUR/TH/15\\
March, 2009\\
\end{flushright}
\vspace{30pt}
\pagestyle{empty}
\baselineskip15pt

\begin{center}
{\large\bf Massive Gravity with Mass Term in Three Dimensions
 \vskip 1mm
}

\vspace{20mm}

Masashi Nakasone
and Ichiro Oda
          \footnote{
           E-mail address:\ ioda@phys.u-ryukyu.ac.jp
                  }

\vspace{10mm}
          Department of Physics, Faculty of Science, University of the 
           Ryukyus,\\
           Nishihara, Okinawa 903-0213, JAPAN \\

\end{center}


\vspace{20mm}
\begin{abstract}
We analyze the effect of the Pauli-Fierz mass term on a recently established, new
massive gravity theory in three space-time dimensions. We show that the Pauli-Fierz mass 
term makes the new massive gravity theory non-unitary. Moreover, although we add 
the gravitational Chern-Simons term to this model, the situation remains unchanged and 
the theory stays non-unitary in spite that structure of the graviton propagator is greatly 
changed.  Thus, the Pauli-Fierz mass term is not allowed to coexist with mass-generating 
higher-derivative terms in the new massive gravity.
\vspace{15mm}

\end{abstract}

\newpage
\pagestyle{plain}
\pagenumbering{arabic}


\rm
\section{Introduction}
It is known that gravitational interaction is naturally coupled to the 
stress-energy tensor just like the electro-magnetic interaction is done to the 
electro-magnetic current. Since the stress-energy tensor is generated
by mass (as well as momentum) of particles, understanding the quantum-mechanical
meaning of mass is a crucial step in developing a quantum theory of gravity. 
Although the graviton mediating the gravitational interaction is a massless 
particle in Einstein's general relativity, there was an old attempt to 
getting the massive graviton by Pauli and Fierz \cite{Fierz}. It seems that 
this attempt of giving mass to the graviton has some applications to 
recent developments of quantum gravity, string theory, 
brane world, and cosmology and so on \cite{Percacci, Kaku1, Porrati, Kirsch, 
't Hooft, Kaku2, Oda1, Maeno1, Maeno2}. In hindsight, this is not surprising
since general relativity is almost the unique theory of massless spin 2 gravitational 
field whose universality class is determined by local general coordinate transformations, 
any infrared modification of general relativity cannot aviod introduction 
of some kind of mass for the graviton.

In an arbitrary space-time dimension, if we do not introduce the other matter fields,
there is the unique mass-generating mechanism for the graviton, which is adding 
the Pauli-Fierz mass term to the Einstein-Hilbert action. This mechanism makes it possible 
to generate mass to the graviton in a Lorentz-covariant manner without the emergence 
of a non-unitary $\it{ghost}$. However, there is at least one serious drawback in 
the Pauli-Fierz massive gravity in that the massive gravity only makes sense 
as a free and linearized theory, so it is unclear how to obtain a diffeomorphism-invariant 
mass and interaction terms in this framework.

Recently, in three space-time dimensions there has been an interesting progress 
for obtaining a sensible interacting massive gravity theory \cite{Bergshoeff}. 
This model has been shown to be equivalent to the Pauli-Fierz massive gravity at the 
linearized approximation level. A key idea in this model is that one adds 
higher-derivative curvature terms to the Einstein-Hilbert action with the $\it{wrong}$ sign 
in such a way that the trace part of the stress-energy tensor associated with those 
higher-derivative terms is proportional to the original higher-derivative Lagrangian. 
More recently, this new massive gravity model in three dimensions has been studied from
various viewpoints such as the unitarity and the impossibility of generalization to higher 
dimensions \cite{Nakasone}, the AdS black hole solutions \cite{Clement} and the 
properties of linearized gravitational excitations in asymptotically AdS space-time
\cite{Liu}.

As a peculiarity of three dimensions, there is an alternative mass-generating mechanism
for the graviton: adding a topological term named the gravitational Chern-Simons term
\cite{Deser}. Thus, in three dimensions, in total there are three distinct mass-generating 
mechanisms for the graviton, those are, adding the Pauli-Fierz, the higher-derivative,
and the gravitational Chern-Simons terms. It is then natural to ask what becomes of mass of the
graviton if these three terms coexist in one theory \footnote{In the case of gauge fields,
the relation between the conventional Higgs mechanism and the topological mass-generating
mechanism was investigated in \cite{Oda2, Oda3}.}. Surprisingly, these seemingly 
innocuous models are not physically acceptable owing to the emergence of ghosts and/or
tachyons. 

The aim of this article is to show that the Pauli-Fierz mass term is not allowed to
exist in the new massive gravity in three dimensions. In this connection, 
it has been already verified that there is a consistent, unitary massive
gravity with three massive excitations in some parameter region between mass 
and the coefficient of gravitational Chern-Simons terms when the Pauli-Fierz term 
is added to the topologically massive gravity in three dimensions \cite{Pinheiro, Tekin}.

In the next section, we clarify why there is an interacting unitary massive
gravity theory only in three dimensions via the analysis of structure of the graviton
propagator in a general space-time dimension. In the third section, we consider the case 
that the Pauli-Fierz mass term is added to the new massive gravity theory and
examine if the original massive graviton stays unitary or not. Here we realize that
the existence of the Pauli-Fierz mass term has a tendency to make the unitary modes
change to ghosts and tachyons. In the final section, we study the most general massive gravity model
where the gravitational Chern-Simons term is added to the model treated in the section 3. 
Again we will see that there appear ghosts in the mass spectrum, so that the Pauli-Fierz
term cannot afford to exist in the new massive gravity theory in three dimensions.

\section{Propagator in a general higher-derivative gravity}

In this section, we wish to pursue a possibility of constructing
a renormalizable, interactive, unitary model with higher-derivative terms
for the massive graviton by studying structure of the graviton propagator. 
Even if it is well known that it is impossible to construct such an ideal 
model in four space-time dimensions \cite{Stelle}, it is useful to clarify 
the meaning of the new massive gravity model by Bergshoeff et al. \cite{Bergshoeff} 
in three dimensions.

The action with which we start is a higher-derivative gravity model \cite{Stelle, Nakasone} 
without cosmological constant up to fourth-order in derivative in a general $D$ space-time 
dimensions \footnote{The space-time indices $\mu, \nu, \cdots$ run over $0, 1, 
\cdots, D-1$. We take the metric signature $(-, +, \cdots, +)$ and follow the notation 
and conventions of the textbook of MTW \cite{MTW}.}:  
\begin{eqnarray}
S = \int d^D x \sqrt{- g} [ \frac{1}{\kappa^2} R  + \alpha R^2
+ \beta R_{\mu\nu} R^{\mu\nu}
+ \gamma( R_{\mu\nu\rho\sigma} R^{\mu\nu\rho\sigma} 
- 4 R_{\mu\nu} R^{\mu\nu}
+ R^2 ) ],
\label{Action1}
\end{eqnarray}
where $\kappa^2 \equiv 16 \pi G_D$ ($G_D$ is the $D$-dimensional Newton's constant), 
$\alpha$, $\beta$ and $\gamma$ are constants. 
The last term proportional to $\gamma$ is nothing but the Gauss-Bonnet term, 
which is a surface term in four space-time dimensions. 

Now let us expand the metric around a flat Minkowski background $\eta_{\mu\nu}$ as 
$g_{\mu\nu} = \eta_{\mu\nu} + h_{\mu\nu}$ and keep only quadratic fluctuations 
in the action. It is convenient to express each term in the action in terms of
the spin projection operators:  
\begin{eqnarray}
{\cal{L}}_{EH} &\equiv& \sqrt{- g} R 
\nonumber\\
&=& \frac{1}{4} h^{\mu\nu} [ P^{(2)} - (D-2) P^{(0, s)} ]_{\mu\nu, \rho\sigma}
\Box h^{\rho\sigma},
\nonumber\\
{\cal{L}}_{\alpha} &\equiv& \alpha \sqrt{- g} R^2 
\nonumber\\
&=& \alpha (D-1) h^{\mu\nu} P^{(0, s)}_{\mu\nu, \rho\sigma}
\Box^2 h^{\rho\sigma},
\nonumber\\
{\cal{L}}_{\beta} &\equiv& \beta \sqrt{- g} R_{\mu\nu} R^{\mu\nu} 
\nonumber\\
&=&  \beta \frac{1}{4} h^{\mu\nu} [ P^{(2)} + D P^{(0, s)} ]_{\mu\nu, \rho\sigma}
\Box^2 h^{\rho\sigma},
\nonumber\\
{\cal{L}}_{\gamma} &\equiv& \gamma \sqrt{- g} ( R_{\mu\nu\rho\sigma} R^{\mu\nu\rho\sigma} 
- 4 R_{\mu\nu} R^{\mu\nu}
+ R^2 )
\nonumber\\
&=& 0,
\label{Lagrangian}
\end{eqnarray}
where in evaluating the last ${\cal{L}}_{\gamma}$ we have used the relation
\begin{eqnarray}
\sqrt{- g} R_{\mu\nu\rho\sigma} R^{\mu\nu\rho\sigma} 
= h^{\mu\nu} [ P^{(2)} + P^{(0, s)} ]_{\mu\nu, \rho\sigma}
\Box^2 h^{\rho\sigma}.
\label{Supple}
\end{eqnarray}

The spin projection operators in $D$ space-time dimensions, 
$P^{(2)}, P^{(1)}, P^{(0, s)}, P^{(0, w)}, P^{(0, sw)}$
and  $P^{(0, ws)}$ 
form a complete set in the space of second rank symmetric tensors
and are defined as \cite{Nakasone}
\begin{eqnarray}
P^{(2)}_{\mu\nu, \rho\sigma} &=& \frac{1}{2} ( \theta_{\mu\rho} \theta_{\nu\sigma}
+ \theta_{\mu\sigma} \theta_{\nu\rho} ) - \frac{1}{D-1} \theta_{\mu\nu} \theta_{\rho\sigma},
\nonumber\\  
P^{(1)}_{\mu\nu, \rho\sigma} &=& \frac{1}{2} ( \theta_{\mu\rho} \omega_{\nu\sigma}
+ \theta_{\mu\sigma} \omega_{\nu\rho} + \theta_{\nu\rho} \omega_{\mu\sigma}
+ \theta_{\nu\sigma} \omega_{\mu\rho} ),
\nonumber\\  
P^{(0, s)}_{\mu\nu, \rho\sigma} &=& \frac{1}{D-1} \theta_{\mu\nu} \theta_{\rho\sigma},
\nonumber\\  
P^{(0, w)}_{\mu\nu, \rho\sigma} &=& \omega_{\mu\nu} \omega_{\rho\sigma},
\nonumber\\  
P^{(0, sw)}_{\mu\nu, \rho\sigma} &=& \frac{1}{\sqrt{D-1}} \theta_{\mu\nu} \omega_{\rho\sigma},
\nonumber\\  
P^{(0, ws)}_{\mu\nu, \rho\sigma} &=& \frac{1}{\sqrt{D-1}} \omega_{\mu\nu} \theta_{\rho\sigma}.
\label{Spin projectors}
\end{eqnarray}
Here the transverse operator $\theta_{\mu\nu}$ and the longitudinal operator
$\omega_{\mu\nu}$ are defined as
\begin{eqnarray}
\theta_{\mu\nu} &=& \eta_{\mu\nu} - \frac{1}{\Box} \partial_\mu \partial_\nu 
= \eta_{\mu\nu} - \omega_{\mu\nu}, \nonumber\\  
\omega_{\mu\nu} &=& \frac{1}{\Box} \partial_\mu \partial_\nu.
\label{theta}
\end{eqnarray}
It is straightforward to show that the spin projection operators satisfy 
the orthogonality relations
\begin{eqnarray}
P_{\mu\nu, \rho\sigma}^{(i, a)} P_{\rho\sigma, \lambda\tau}^{(j, b)}
&=& \delta^{ij} \delta^{ab} P_{\mu\nu, \lambda\tau}^{(i, a)},
\nonumber\\  
P_{\mu\nu, \rho\sigma}^{(i, ab)} P_{\rho\sigma, \lambda\tau}^{(j, cd)}
&=& \delta^{ij} \delta^{bc} P_{\mu\nu, \lambda\tau}^{(i, a)},
\nonumber\\  
P_{\mu\nu, \rho\sigma}^{(i, a)} P_{\rho\sigma, \lambda\tau}^{(j, bc)}
&=& \delta^{ij} \delta^{ab} P_{\mu\nu, \lambda\tau}^{(i, ac)},
\nonumber\\  
P_{\mu\nu, \rho\sigma}^{(i, ab)} P_{\rho\sigma, \lambda\tau}^{(j, c)}
&=& \delta^{ij} \delta^{bc} P_{\mu\nu, \lambda\tau}^{(i, ac)},
\label{Orthogonality}
\end{eqnarray}
with $i, j = 0, 1, 2$ and $a, b, c, d = s, w$ and the tensorial relation 
\begin{eqnarray}
[ P^{(2)} + P^{(1)} + P^{(0, s)} + P^{(0, w)} ]_{\mu\nu, \rho\sigma} = 
\frac{1}{2} ( \eta_{\mu\rho}\eta_{\nu\sigma} + \eta_{\mu\sigma}\eta_{\nu\rho} ).
\label{T relation}
\end{eqnarray}

Using the relations (\ref{Lagrangian}), the action (\ref{Action1}) takes the
form
\begin{eqnarray}
S = \int d^D x \frac{1}{4} h^{\mu\nu} {\cal{O}}_{\mu\nu, \rho\sigma} 
h^{\rho\sigma},
\label{Action2}
\end{eqnarray}
where ${\cal{O}}_{\mu\nu, \rho\sigma}$ is defined as
\begin{eqnarray}
{\cal{O}}_{\mu\nu, \rho\sigma} = \Box [ ( \frac{1}{\kappa^2} + \beta \Box ) P^{(2)} 
+ \{ - \frac{1}{\kappa^2}(D-2) + ( 4 \alpha (D-1) + \beta D ) \Box \} 
P^{(0, s)} ]_{\mu\nu, \rho\sigma}.
\label{O}
\end{eqnarray}
Then, the graviton propagator is essentially obtained by inverting each spin block
\begin{eqnarray}
&{}& {\cal{O}}_{\mu\nu, \rho\sigma}^{-1} = \frac{1}{\Box}
[ \frac{1}{\frac{1}{\kappa^2} + \beta \Box} P^{(2)} 
+  \frac{1}{- \frac{1}{\kappa^2} (D-2) + ( 4 \alpha
(D-1) + \beta D) \Box} P^{(0, s)} ]_{\mu\nu, \rho\sigma}
\nonumber\\
&=& \kappa^2 [ \frac{P^{(2)} - \frac{1}{D-2} P^{(0, s)}} {\Box}  
- \frac{1}{\Box + \frac{1}{\beta\kappa^2}} P^{(2)}
+ \frac{1}{D-2} \frac{1}{\Box - \frac{D-2}{ \kappa^2 (4 \alpha (D-1) + \beta D)}} 
P^{(0, s)} ]_{\mu\nu, \rho\sigma}.
\label{Inverse O}
\end{eqnarray}
Recall that the combination $P^{(2)} - \frac{1}{D-2} P^{(0, s)}$ is 
parallel to that in four dimensions \cite{Nieu} and the negative sign
for the massless pole with $P^{(0, s)}$ does not impose any problem \cite{Nakasone}.

This structure of the graviton propagator clearly indicates why
it is difficult to construct a unitary gravitational theory for the massive graviton
in a general space-time dimension. One point which we immediately notice is that
the negative sign for the massive pole $\Box = - \frac{1}{\beta\kappa^2}$
corresponds to a ghost with negative norm, so in order to avoid this
massive ghost, we have to impose the condition $\beta = 0$, implying that 
$R_{\mu\nu}^2$ term is not allowed to be involved in the action up
to the Gauss-Bonnet term. 
On the other hand, the remaining two poles, massless and
massive poles, correspond to spin 2 and 0 unitary modes. (Of course,
we have to impose the further conditon $\alpha > 0$ in order to avoid
tachyons.) In fact, with the vanishing $R_{\mu\nu}^2$ term, 
up to the Gauss-Bonnet term, the action (\ref{Action1}) is reduced to $R + R^2$, 
which is known to exactly coincide with Einstein's general relativity with a
minimally coupled massive scalar field \cite{Schmidt}. 
In this way, it turns out to be difficult
to construct a unitary massive gravity model with helicity $\pm 2$ massive graviton 
modes within the framework of the higher-derivative gravity at least in an arbitrary
space-time dimension.

However, by inspection, we notice that if the the gravitational coupling constant 
$\kappa^2$ $\it{were}$ negative, the residue at the pole 
$\Box = - \frac{1}{\beta\kappa^2}$ would become positive whereas those at the massless 
$\Box = 0$ and massive poles $\Box = \frac{D-2}{ \kappa^2 (4 \alpha (D-1) + \beta D)}$ 
do negative, thereby implying that the modes corresponding to the former pole and the 
latter two poles are respectively unitary modes with positive norm and ghosts with 
negative norm. At this stage, a nice thing happens in three dimensions. Namely,
in three dimensions, it turns out that massless graviton modes are non-dynamical \cite{Nakasone}
so we can neglect the massless graviton whatever its norm is positive or negative.
The remaining problem is therefore cast to a problem how to deal with the massive
scalar mode with negative norm. This problem is overcome by selecting the constants
$\alpha, \beta$ in such a way that they satisfy the relation
\begin{eqnarray}
(4 \alpha (D-1) + \beta D)|_{D=3} = 8 \alpha + 3 \beta = 0. 
\label{Alpha}
\end{eqnarray}
After all, we arrive at a unitary massive gravity action in three dimensions
which has been recently found by Bergshoeff et al. \cite{Bergshoeff}
\begin{eqnarray}
S = \int d^3 x \frac{1}{\kappa^2} \sqrt{- g} [ - R  
+ \frac{1}{M^2} ( R_{\mu\nu} R^{\mu\nu} - \frac{3}{8} R^2 ) ],
\label{Action3}
\end{eqnarray}
where we have set $\beta \equiv \frac{1}{\kappa^2 M^2}$ because of $\beta > 0$.
Note that we have replaced $\kappa^2$ in (\ref{Action1}) with $- \kappa^2$
in (\ref{Action3}) in order to make the sign for the Einstein-Hilbert
term negative.

\section{Massive gravity with the Pauli-Fierz mass term}

Now we shall add the Pauli-Fierz mass term to the new massive gravity 
in three dimensions and examine the effect on the graviton propagator.

For this purpose, let us begin with the action
\begin{eqnarray}
S = \int d^3 x [ \frac{1}{\kappa^2} \sqrt{- g} \{ - R  
+ \frac{1}{M^2} ( R_{\mu\nu} R^{\mu\nu} - \frac{3}{8} R^2 ) \}
- \frac{m^2}{4} ( h_{\mu\nu} h^{\mu\nu} - h^2 ) ],
\label{Action4}
\end{eqnarray}
where $h \equiv \eta^{\mu\nu} h_{\mu\nu}$ and we shall set $\kappa^2 = 1$
henceforth. First, let us note that the Pauli-Fierz mass term can 
be rewritten by the spin projection operators as
\begin{eqnarray}
{\cal{L}}_{PF} &\equiv& 
- \frac{m^2}{4} ( h_{\mu\nu} h^{\mu\nu} - h^2 ) 
\nonumber\\
&=& \frac{m^2}{2} h^{\mu\nu} [ - \frac{1}{2} P^{(2)} - \frac{1}{2} P^{(1)}
+ \frac{1}{2} P^{(0, s)} 
+ \frac{1}{\sqrt{2}} (P^{(0, sw)} + P^{(0, ws)}) ]_{\mu\nu, \rho\sigma} 
h^{\rho\sigma}.
\label{Pauli}
\end{eqnarray}
Thus, the quadratic part of the action (\ref{Action4}) is expressed
in term of the spin projection operators like
\begin{eqnarray}
S = \int d^3 x \frac{1}{2} h^{\mu\nu} {\cal{P}}_{\mu\nu, \rho\sigma} 
h^{\rho\sigma},
\label{Action5}
\end{eqnarray}
where ${\cal{P}}_{\mu\nu, \rho\sigma}$ is defined as
\begin{eqnarray}
{\cal{P}}_{\mu\nu, \rho\sigma} &=& [ \frac{1}{2} ( \frac{1}{M^2} \Box^2
- \Box - m^2 ) P^{(2)} - \frac{m^2}{2} P^{(1)} 
+ \frac{1}{2} ( \Box + m^2 ) P^{(0, s)}
\nonumber\\
&+& \frac{m^2}{\sqrt{2}} ( P^{(0, sw)} + P^{(0, ws)} )]_{\mu\nu, \rho\sigma}.
\label{P}
\end{eqnarray}

Then, the propagator for $h_{\mu\nu}$ is defined by
\begin{eqnarray}
<0| T (h_{\mu\nu}(x) h_{\rho\sigma}(y)) |0>
= i {\cal{P}}_{\mu\nu, \rho\sigma}^{-1} \delta^{(D)}(x-y),
\label{Propa}
\end{eqnarray}
where using the relation (\ref{T relation}), the inverse of the operator $\cal{P}$ 
is easily calculated as
\begin{eqnarray}
{\cal{P}}_{\mu\nu, \rho\sigma}^{-1} 
&=& [ \frac{2}{\frac{1}{M^2} \Box^2 - \Box - m^2} P^{(2)} 
-  \frac{2}{m^2} P^{(1)} 
- \frac{\Box + m^2}{m^4} P^{(0, w)} 
\nonumber\\
&+& \frac{\sqrt{2}}{m^2} ( P^{(0, sw)} + P^{(0, ws)} )]_{\mu\nu, \rho\sigma}.
\label{P-inv}
\end{eqnarray}
The expression of the propagator (\ref{P-inv}) reveals that there are
massive poles in the sector of spin 2 graviton modes, which is of form  
\begin{eqnarray}
I &\equiv& \frac{1}{M^2} \Box^2 - \Box - m^2
\nonumber\\
&=& \frac{1}{M^2} ( \Box - \omega_+ ) ( \Box - \omega_- ),
\label{Pole1}
\end{eqnarray}
where $\omega_{\pm} \equiv \frac{1 \pm \sqrt{1 + 4 (\frac{m}{M})^2}}
{2} M^2$, which are real numbers such that $\omega_+ > 0, \omega_- < 0$.
In order to understand the physical property of the poles, it is useful to rewrite 
$\frac{1}{I}$ as
\begin{eqnarray}
\frac{1}{I} = \frac{1}{\sqrt{1 + 4 (\frac{m}{M})^2}} 
( \frac{1}{\Box - \omega_+} - \frac{1}{\Box - \omega_-} ),
\label{Pole2}
\end{eqnarray}
which shows that there are two massive spin 2 modes. One is the unitary mode 
of positive mass $\omega_+$ with positive norm while the other is
the ghost of tachyonic mass $\omega_-$ with negative norm. Thus in the model
at hand the existence of this tachyonic ghost induces the violation of both unitarity 
and causality, so that adding the Pauli-Fierz mass term to the new massive 
gravity theory in three dimensions is not permitted from the physical 
requirements of unitarity and causality.

\section{Incorporation of 'gravitational' Chern-Simons term}

Next, in this section, we wish to consider the most general situation where three 
distinct mass-generating terms, those are, the Pauli-Fierz, higher-derivative, 
and 'gravitational' Chern-Simons terms, coexist with the Einstein-Hilbert
action in three dimensions. In particular, we are interested in the question 
whether or not the presence of the gravitational Chern-Simons term could 
rectify the impossibility of adding the Pauli-Fierz mass term to the new massive
gravity model. Actually, it has been already pointed out that in the case of 
the topological massive gravity with the Pauli-Fierz mass term, there is a 
parameter region where ghosts and tachyons are simultaneously excluded 
and we have three massive (not tachyonic) excitations with positive norm 
\cite{Pinheiro, Tekin}.

The most general action which includes three mass-generating terms
is given by
\begin{eqnarray}
S &=& \int d^3 x [ \sqrt{- g} \{ - R  
+ \frac{1}{M^2} ( R_{\mu\nu} R^{\mu\nu} - \frac{3}{8} R^2 ) \}
- \frac{m^2}{4} ( h_{\mu\nu} h^{\mu\nu} - h^2 ) 
\nonumber\\
&+& \frac{1}{\mu} \varepsilon^{\mu\nu\rho} \Gamma_{\mu\lambda}^\sigma
( \partial_\nu \Gamma_{\sigma\rho}^\lambda 
+ \frac{2}{3} \Gamma_{\nu\tau}^\lambda \Gamma_{\rho\sigma}^\tau ) ],
\label{Action6}
\end{eqnarray}
where the coefficient $\mu$ in front of the gravitational Chern-Simons term
is a constant.  In order to accommodate with this topological term,
one has to introduce additional two operators to the whole spin projection
operators \cite{Pinheiro}
\begin{eqnarray}
S_{1 \mu\nu, \rho\sigma} &=& \frac{1}{4} 
\Box ( \varepsilon_{\mu\rho\lambda} \partial_\sigma \omega^\lambda_\nu
+ \varepsilon_{\mu\sigma\lambda} \partial_\rho \omega^\lambda_\nu
+ \varepsilon_{\nu\rho\lambda} \partial_\sigma \omega^\lambda_\mu
+ \varepsilon_{\nu\sigma\lambda} \partial_\rho \omega^\lambda_\mu ),
\nonumber\\  
S_{2 \mu\nu, \rho\sigma} &=& - \frac{1}{4} 
\Box ( \varepsilon_{\mu\rho\lambda} \eta_{\sigma\nu}
+ \varepsilon_{\mu\sigma\lambda} \eta_{\rho\nu}
+ \varepsilon_{\nu\rho\lambda} \eta_{\sigma\mu}
+ \varepsilon_{\nu\sigma\lambda} \eta_{\rho\mu} ) \partial^\lambda.
\label{Spin projectors2}
\end{eqnarray}
These operators together with the spin projection operators satisfy the
following relations
\begin{eqnarray}
S_1 S_1 &=& \frac{1}{4} \Box^3 P^{(1)}, \nonumber\\
S_1 S_2 &=& S_2 S_1 = - \frac{1}{4} \Box^3 P^{(1)}, \nonumber\\
S_2 S_2 &=& \Box^3 ( P^{(2)} + \frac{1}{4} P^{(1)} ), \nonumber\\
P^{(1)} S_1 &=& S_1 P^{(1)} = S_1, \nonumber\\
P^{(1)} S_2 &=& S_2 P^{(1)} = - S_1, \nonumber\\
P^{(2)} S_2 &=& S_2 P^{(2)} = S_1 + S_2,
\label{Spin projectors3}
\end{eqnarray}
where the matrix indices are to be understood.

As before, we write out the quadratic fluctuations in $h_{\mu\nu}$ in the action
(\ref{Action6}) whose result reads
\begin{eqnarray}
S = \int d^3 x \frac{1}{2} h^{\mu\nu} {\cal{Q}}_{\mu\nu, \rho\sigma} 
h^{\rho\sigma},
\label{Action7}
\end{eqnarray}
where ${\cal{Q}}_{\mu\nu, \rho\sigma}$ is defined as
\begin{eqnarray}
{\cal{Q}}_{\mu\nu, \rho\sigma} &=& [ \frac{1}{2} ( \frac{1}{M^2} \Box^2
- \Box - m^2 ) P^{(2)} - \frac{m^2}{2} P^{(1)} 
+ \frac{1}{2} ( \Box + m^2 ) P^{(0, s)}
\nonumber\\
&+& \frac{m^2}{\sqrt{2}} ( P^{(0, sw)} + P^{(0, ws)} )
+ \frac{1}{\mu} ( S_1 + S_2 ) ]_{\mu\nu, \rho\sigma}.
\label{Q}
\end{eqnarray}
The inverse matrix of ${\cal{Q}}_{\mu\nu, \rho\sigma}$, which is proportional
to the graviton propagator, is calculated as 
\begin{eqnarray}
{\cal{Q}}_{\mu\nu, \rho\sigma}^{-1} 
&=& [ \frac{2 (\frac{1}{M^2} \Box^2 - \Box - m^2)}
{(\frac{1}{M^2} \Box^2 - \Box - m^2)^2 - \frac{4}{\mu^2} \Box^3} P^{(2)} 
-  \frac{2}{m^2} P^{(1)} 
- \frac{\Box + m^2}{m^4} P^{(0, w)} 
\nonumber\\
&+& \frac{\sqrt{2}}{m^2} ( P^{(0, sw)} + P^{(0, ws)} )
- \frac{4}{\mu} \frac{1}{(\frac{1}{M^2} \Box^2 - \Box - m^2)^2 
- \frac{4}{\mu^2} \Box^3} ( S_1 + S_2 ) ]_{\mu\nu, \rho\sigma}.
\label{Q-inv}
\end{eqnarray}
It is easy to see that this expression reduces to (\ref{P-inv})
in the limit of $\mu \rightarrow \infty$ as required.

As opposed to the previous case without the gravitational Chern-Simons term
(\ref{P-inv}), there appears quartic pole in the sectors of $P^{(2)}, S_1$ and $S_2$. 
Thus, there might be a certain parameter region where a unitary and tachyon-free
massive gravity theory exists. In order to examine this possibility,
let us consider the pole structure 
\begin{eqnarray}
J &\equiv& (\frac{1}{M^2} \Box^2 - \Box - m^2)^2 - \frac{4}{\mu^2} \Box^3
\nonumber\\
&=& (\frac{1}{M^2} \Box^2 - \Box - m^2 + \frac{2}{\mu} \Box^{\frac{3}{2}}) 
(\frac{1}{M^2} \Box^2 - \Box - m^2 - \frac{2}{\mu} \Box^{\frac{3}{2}}).
\label{Pole3}
\end{eqnarray}
The necessary condition that there is no ghost is given by the condition
such that the quartic equation $J = 0$ has no real solution. It then turns 
out that the equation $J = 0$ has indeed real solutions for any real value
of $M, m$ and $\mu$. Hence, even in this case where the gravitational
Chern-Simons term is added to the Pauli-Fierz term plus the new massive gravity
theory, there is no physically plausible, unitary massive gravity theory.
In this sense, the Pauli-Fierz mass term is not allowed to exist in the
new massive gravity model in three dimensions. 
In this context, let us recall that the Pauli-Fierz massive
gravity makes sense only as a free theory holding in the quadratic approximation 
level in $h_{\mu\nu}$ so the theory is not diffeomorphism-invariant but 
Lorentz-invariant. On the other hand, the new massive gravity is 
an interacting and diffeomorphism-invariant theory, so the theory
might not admit the existence of the diffeomorphism-noninvariant
Pauli-Fierz term.

\vs 1   

\end{document}